\documentclass[%
 reprint,
 amsmath,amssymb,
 aps,
 prl,
]{revtex4-2}

\newcommand{\RNum}[1]{\uppercase\expandafter{\romannumeral #1\relax}}
\usepackage{siunitx}
\usepackage{graphicx}% Include figure files
\usepackage{dcolumn}% Align table columns on decimal point
\usepackage{bm}% bold math
\usepackage{xcolor}
\usepackage{array}
\usepackage{multirow}

\usepackage{mathtools}

\DeclarePairedDelimiter{\abs}{\lvert}{\rvert}

\begin{document}

\preprint{APS/123-QED}

\title{%
Impact of transduction scaling laws on nanoelectromechanical systems
}%

\author{Konstantinos Tsoukalas}\email{ktso@dtu.dk}
\author{Babak Vosoughi Lahijani}
\author{S{\o}ren Stobbe}

\affiliation{%
Department of Photonics Engineering, DTU Fotonik, Technical University of Denmark, Building 343, DK-2800 Kgs.\ Lyngby, Denmark
}%

\date{\today}% It is always \today, today,
%https://www.overleaf.com/5985446199wrnrnnthhstq
%https://www.overleaf.com/6443786525zjvpngtnnnsj

\begin{abstract}
We study the electromechanical transduction in nanoelectromechanical actuators and show that the differences in scaling for electrical and mechanical effects lead to an overall non-trivial scaling behavior. In particular, the previously neglected fringing fields considerably increase electrical forces and improve the stability of nanoscale actuators. This shows that electrostatics does not pose any limitations to downscaling of electromechanical systems, in fact in several respects, nanosystems outperform their microscale counterparts. As a specific example, we consider in-plane actuation of ultrathin slabs and show that devices consisting of a few layers of graphene are feasible, implying that electromechanical resonators operating beyond 40 GHz are possible with currently available technology.
\end{abstract}

\keywords{electrostatic actuation, aspect ratio, fringing fields, two-dimensional actuation, comb-drive, co-planar strips, parallel beams, electrostatic instability, pull-in}%Use showkeys class option if keyword
\maketitle

%\tableofcontents

\begin{figure}[t]
\includegraphics[width=\columnwidth]{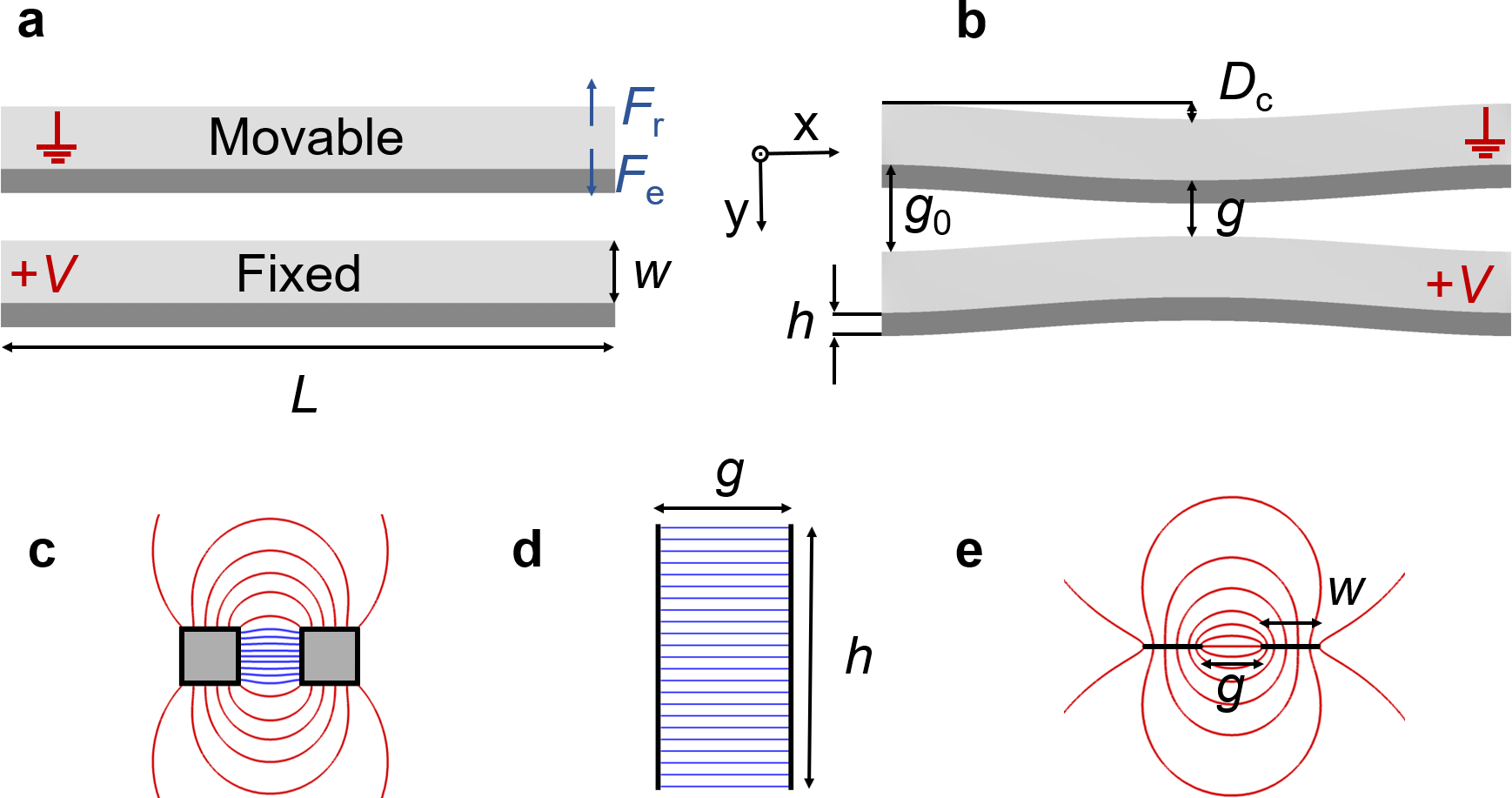}
\includegraphics[width=\columnwidth]{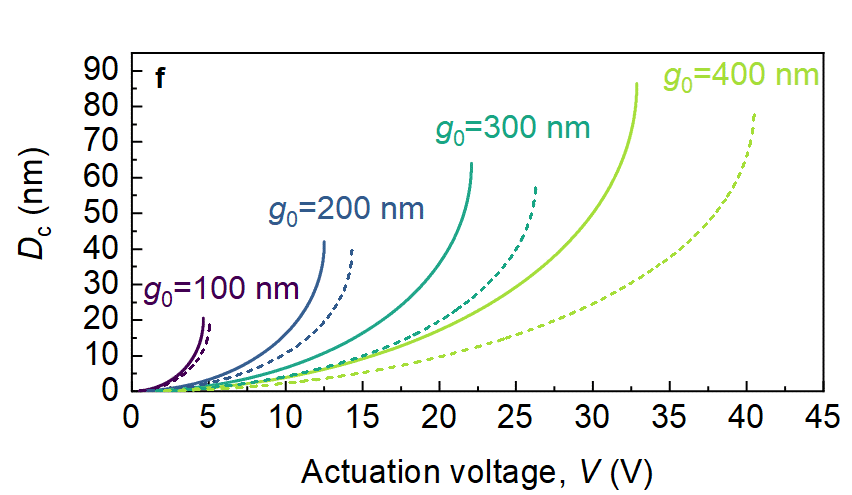}
\caption{\label{fig:1}Geometries and electric fields of NEMS electrostatic actuators. Electrostatic actuators consisting of (\textbf{a}) rigid and (\textbf{b}) elastic beams exhibit (\textbf{c}) electric-field lines, which combine the characteristics of the fields of a (\textbf{d}) parallel-plate and (\textbf{e}) co-planar strips. (\textbf{f}) The displacement of elastic beams for experimentally relevant parameters calculated with the full electric field (solid) is significantly larger than in the parallel-plate approximation (dashed).
}
\end{figure}

Microelectromechanical systems (MEMS) are prevalent in modern microtechnologies in the form of microphones, gyroscopes, accelerometers, magnetometers, thermometers, as well as pressure sensors and chemical sensors \cite{Madou,Senturia}. A most central MEMS functionality is the transduction between electromagnetic and mechanical degrees of freedom, the simplest of which relies on electrostatics where the electrostatic attraction between mechanically compliant parts results in reversible elastic deformation \cite{TANG1990,Nguyen2007,pelesko,Younis2003,Osterberg97}. Electrostatic actuators have found applications ranging from experimental investigations of fundamental physics \cite{casimirexp} over micro- and nanorobotics \cite{robotics} to nanophotonics \cite{actReviw,Midolo2018}, and they were recently proposed as building blocks for adiabatic computing \cite{computation}.

A current research frontier is now concerned with scaling MEMS to smaller dimensions, i.e., developing nanoelectromechanical systems (NEMS) \cite{Roukes2001} and the physics of NEMS devices is often described via scaling arguments based on the parallel-plate capacitor model \cite{Madou,Senturia,Baglio08}. This down-scaling is desirable not only to reduce the footprint and the consumption of materials and energy, but also to enable new functionalities. For example, electrochemistry and fluidics set the length scale of electromechanical integration with biological systems \cite{biomems} and NEMS combined with nanophotonics opens a wealth of opportunities because of the subwavelength nature of the mechanical components \cite{Midolo2018}. However, NEMS is not merely scaled-down MEMS because atomic and quantum effects, such as the Casimir effect \cite{casimir,Fong2019}, field emission \cite{fieldem}, and size-dependence of the Young's modulus \cite{young} become significant at the nanoscale. In addition, constraints imposed by applications as well as limitations to nanofabrication technology imply that aspect ratios \cite{asprate,vic1,Akarvardar} and more generally the design features in NEMS are different from their MEMS counterparts. For example, in NEMS photonics, these constraints in practice favor an aspect ratio close to unity \cite{Midolo2018,cami,winger,Akihama2012,vic4,vic3,spectrometer}, which diminishes the transduction calculated in the parallel-plate model \cite{badscale,Pfeifer16}. This calls for a theoretical investigation of electromechanical actuation for arbitrary aspect ratios in MEMS and NEMS, which was so far missing. Here we consider the impact of scaling on the transduction in NEMS actuators for general aspect ratios as shown in Figs.~\ref{fig:1}\textbf{a}-\textbf{e}. We find that the fringing fields significantly change the scaling laws, instabilities, and actuation range.

We first consider an electrostatic actuator consisting of two rectangular beams (Figs.~\ref{fig:1}\textbf{a},\textbf{b}). The cross-section electric-field lines for a finite aspect ratio $h/w$, where $h$ is the height and $w$ is the width of the beams, is illustrated in Fig.~\ref{fig:1}\textbf{c}. The field consists of a combination of the characteristics of the two limiting cases of $h \gg w$, which is a parallel-plate capacitor with a homogeneous field as shown in Fig.~\ref{fig:1}\textbf{d},
and $h \ll w$, which are co-planar strips with a highly inhomogeneous fringing field as shown in Fig.~\ref{fig:1}\textbf{e}.

The displacement, $D_\text{c}$, at the center of two elastic beams (Fig.~\ref{fig:1}\textbf{b}) can be calculated from the Euler-Bernoulli equations with fixed endpoints and subject to an electrostatic pressure \cite{Younis2003,beamfringe}. The latter can be calculated from the capacitance per unit length, $C_\text{2D}$, which we derive from the capacitance of a rectangle near a ground plane \cite{Chuang12,SM},
 \begin{equation}
\label{eq:C2da}
   C_\text{2D} =  \frac{\epsilon_0}{2} \Bigg( \frac{2h}{g} + c \bigg( \frac{2w}{g} \bigg)^{a} + d \bigg( \frac{h}{w} \bigg) ^{b} +f \Bigg),
\end{equation}
where $\epsilon_0$ is the vacuum permittivity, $g$ the the cross-sectional distance, $a=b=0.23$, $c = 3.31$, $d =0.73$, $f = -1.06$ and is formally valid for $h/g \geq 0.05$ and $0.05 \leq w/g \leq 5$, but we find that it agrees with detailed numerical models even beyond these ranges (See Appendix \ref{A1}). The first term in Eq.~(\ref{eq:C2da}) corresponds to a parallel-plate capacitor, the second describes the fringing fields, and the other two do not contribute to the force in this geometry. To illustrate the impact of the fringing fields, we consider silicon beams with width $w= \SI{230}{\nm}$, height $h= \SI{220}{\nm}$, and length $L = \SI{15}{\um}$, and Young's modulus $E= \SI{169}{\giga \pascal}$, which are experimentally realistic parameters for photonic NEMS \cite{cami,spectrometer,srin}. The result for different initial gaps, $g_\text{0}$, is shown in Fig.~\ref{fig:1}\textbf{f}, which also shows the result obtained without the fringing fields. Notably, the deformation is significantly higher than predicted by the inaccurate yet commonly employed parallel-plate approximation, e.g., it is approx.\ two times higher for $g_0 = \SI{100}{\nm}$ and a voltage of $V = \SI{4.65}{\volt}$.

The maximum achievable displacement for electrostatic actuators is only a fraction of the initial gap due the pull-in instability \cite{pelesko,ZHANG2014187}. It occurs when the applied voltage is increased beyond a critical value, $V_\text{p}$, beyond which the mechanical restoring force cannot counterbalance the electrostatic force and the two bodies collapse onto each other. More generally, any force, that obeys an inverse power-law,
\begin{equation}
\label{eq:power}
    F\propto\frac{1}{(g_0-y)^\nu},
\end{equation}
and is balanced by an elastic restoring force, has a maximum actuation range of $g_0/ (\nu +1)$. For example, a parallel-plate actuator has $\nu = 2$ and is therefore limited to displacements of $1/3$ the initial gap. An important finding of our work, which is already apparent from Fig.~\ref{fig:1}\textbf{f}, is that the fringing fields increase the actuation range beyond the parallel-plate instability point $g_0/3$. As we will show in the following, this advantage of low-aspect-ratio NEMS originates from the scaling of the electrostatic force. 

\begin{figure}
\includegraphics[width=\columnwidth]{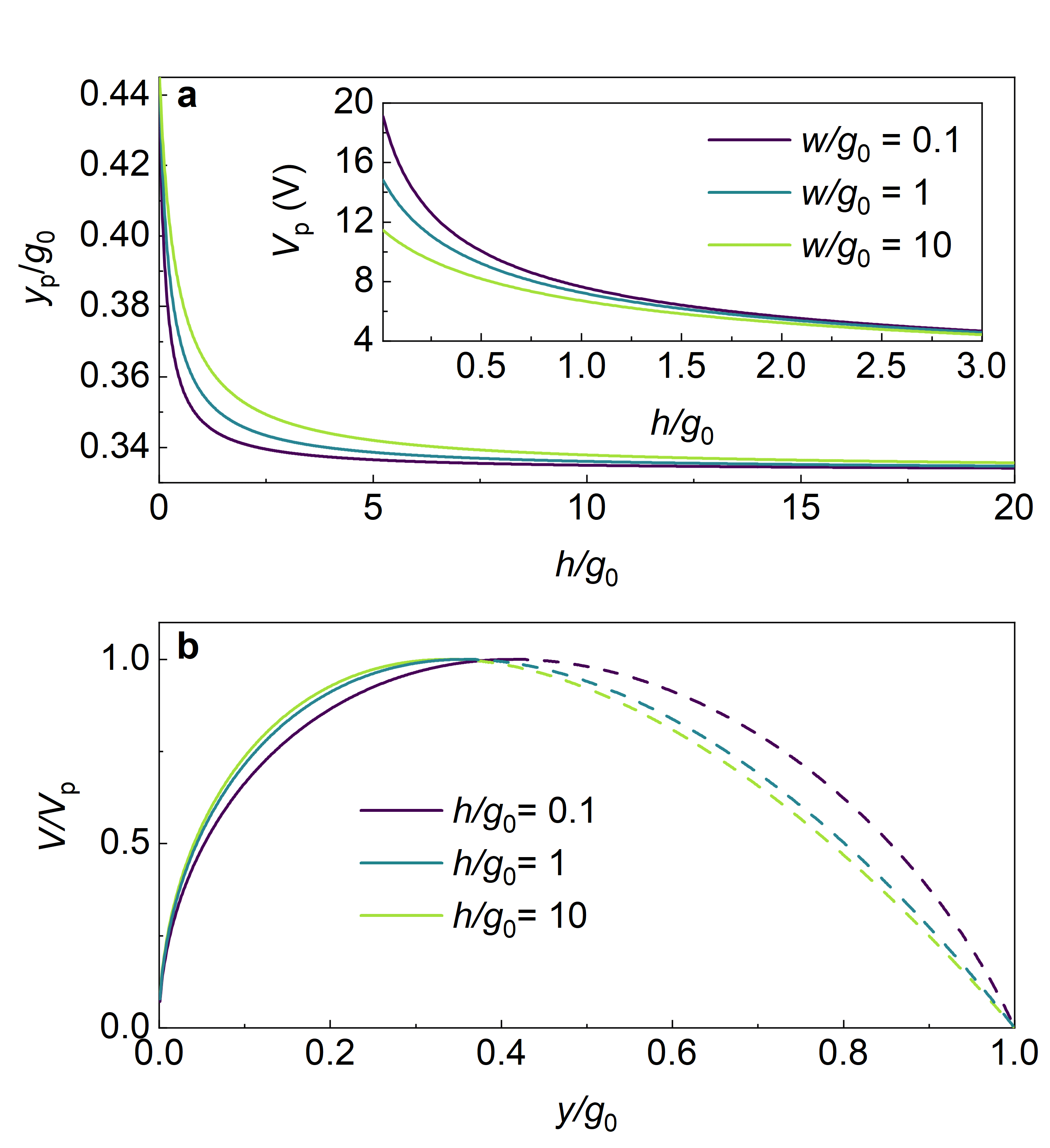}
\caption{\label{fig:beams} Displacement and instability in parallel-beams transduction. (\textbf{a}) Normalised pull-in actuation range of a rigid parallel-beam actuator (Fig.~\ref{fig:1}\textbf{a}) for different ratios of the height to the initial gap. The parallel-plate approximation is valid for electrostatic actuators with high aspect ratios as the normalized pull-in distance $y_\text{p}/g_0$ approaches $1/3$ for large beam heights. Inset: Pull-in voltage for different height-to-gap and width-to-gap ratios. The spring constant is assumed $k=\SI{0.5}{\newton \meter ^{-1}}$ for all widths and heights. The beam width has a significant contribution to the pull-in voltage. (\textbf{b}) Stable (solid) and unstable (dashed) equilibrium curves for different ratios of height to initial gap, $h/g_0$, and $w=g_0$.}
\end{figure}
 
In order to study the general behaviour of electrostatic actuation, we use a simple lumped spring-mass model comprised of two rigid parallel beams, where one beam is fixed and the other one is attached to a spring (Fig.~\ref{fig:1}\textbf{a}). The total electromechanical force on the movable beam is 
(See Appendix)
\begin{equation}
    \label{eq:ftot}
    F =  \frac{\epsilon_\text{0} L}{2} \Bigg( \frac{h}{(g_0-y)^2} +  \frac{2^{a-1} w^a c a}{(g_0-y)^{a+1}}  \Bigg) V^2 - ky.
\end{equation}
Figure~\ref{fig:beams}\textbf{a} shows the normalized pull-in displacement as a function of the ratio of height to initial gap, for three different ratios of the width to the initial gap, calculated from Eq.~(\ref{eq:ftot}).
We observe that the pull-in distances lie between those of the limiting cases of $h \gg w$ $(y_\text{p}=g_0/3)$ and $h \ll w$  $(y_\text{p}=g_0/2.23)$ in accordance with Eq.~(\ref{eq:power}).
In addition, the pull-in voltage is much larger for flat actuators ($h \ll g_0$) as shown in the inset of Fig.~\ref{fig:beams}\textbf{a}. We note that the pull-in voltage depends on the spring constant $k$, which in actual devices may likely depend on $h$ and $w$ but is kept constant in Fig.~\ref{fig:beams}\textbf{a} to enable a direct comparison. In contrast, the pull-in displacement and normalized voltage (shown in Figure~\ref{fig:beams}\textbf{b}) are independent of the spring constant (See Appendix).
Notwithstanding the importance of the increased pull-in displacement for flat actuators, the most striking conclusion from Fig.~\ref{fig:beams} is that the quantity of most immediate experimental importance, namely the voltage-displacement relation (Fig.~\ref{fig:beams}\textbf{b}) is only weakly dependent on the aspect ratio.
This indicates that very thin electrostatic actuators are feasible thanks to the fringing fields taking over the actuation from the homogeneous fields.

\begin{figure}
\includegraphics[width=\columnwidth]{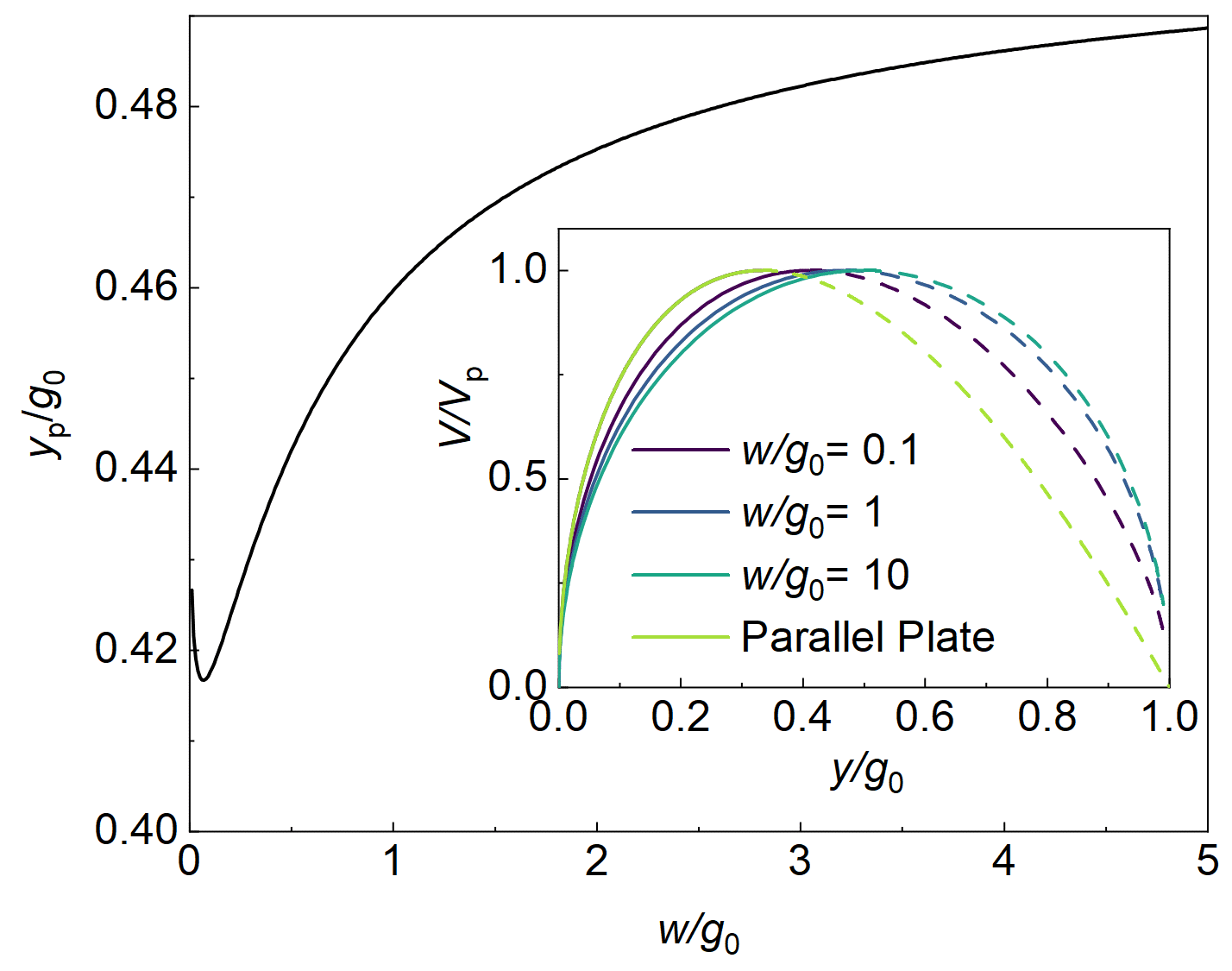}
\caption{\label{fig:microstrip} Normalised pull-in actuation range of a co-planar strip actuator for different ratios of width to initial gap. The system corresponds to a parallel-beam actuator of zero height with a cross-section shown in Figure \ref{fig:1}\textbf{f}. Inset: stable (solid) and unstable (dashed) equilibrium curves for different strip width to initial gap ratio. The actuation range is greater than that of a parallel-plate actuator due to the nature of the fringing forces.}
\end{figure}

Motivated by the apparent feasibility of actuators with ultra-small aspect ratio, we investigate this further using the lumped spring-mass model for the limiting case of two co-planar strips. This model can describe ultra-thin membranes \cite{ultrathin} and two-dimensional materials \cite{suspended} for which in-plane electrical transduction has not been investigated before. The capacitance per unit length of two co-planar strips is given analytically by $C_s = \epsilon_0 K(\sqrt{1-q^2})/K(q)$ \cite{strips}, where $q = g/(2w+g)$ and $K$ is the complete elliptic integral of the first kind. The normalised pull-in displacement as a function of width-to-initial gap ratio is calculated numerically (See Appendix) and shown in Fig.~\ref{fig:microstrip}. The actuation range is always more than $40\;\%$ of the initial gap and we prove that it asymptotically reaches $1/2$ (See Appendix), showing that actuation of 2D beams outperforms those of finite height. The actuation range increases with increasing width but behaves non-monotonically for very small width-to-gap ratios. This is explained by the non-trivial dependence of the capacitance on the width, which in this case, but also in other 2D geometries, is asymptotically logarithmic (See Appendix). Finally, the voltage-displacement dependence (inset in Fig.~\ref{fig:microstrip}) shows a qualitative behaviour of the co-planar strip actuator that is similar to that of the parallel-plate actuator.

As a concrete example of in-plane actuation mechanisms, we consider two parallel co-planar strips of a suspended graphene monolayer with $w= \SI{450}{\nm}$, $L= \SI{7}{\um}$ and $g_0 = \SI{100}{\nm}$. The mechanical properties of graphene \cite{graphene} are characterised by the Young's modulus $E= \SI{1}{\tera \pascal}$, density $\rho = \SI{2200}{\kg \m ^{-3}}$ and the sheet height $h= \SI{0.35}{\nm}$. For a doubly clamped beam, the lumped parameter $k$ is given by $ k= 32Eh (w/L)^3$, which leads to a spring constant of $k = \SI{2.98}{\newton \meter ^{-1}}$. The maximum actuation can be found from Fig.~\ref{fig:microstrip} to be $x_\text{p} = \SI{49}{\nm}$ and the pull-in voltage is calculated to be $V_\text{p}=\SI{20}{\volt}$ (See Appendix). The fundamental frequency of a beam with a negligible built-in-tension is given by $f_0 = 1.03 w \sqrt{E/\rho} /L^2$, which predicts an actuation frequency of $\SI{200}{\mega \hertz}$. These numbers suggest that in-plane actuation can be used for monolithic graphene NEMS that can take full advantage of the material properties and represent the ultimate limit of 2D devices \cite{Bachtold08} although previous research has focused on out-of-plane actuation. For given beam dimensions, the in-plane to out-of-plane resonance frequency scales as $f_\text{in}/f_\text{out}=w/h$, indicating orders of magnitude difference in oscillations of 2D materials. The highest frequency demonstrated via out-of-plane actuation is $f_\text{out}=\SI{1.17}{\giga \hertz}$ \cite{Jung19} for a beam of $w= \SI{3}{\um}$, $L= \SI{1.2}{\um}$ and has been achieved by taking advantage of a very high built-in tension, but the out-of-plane frequency of this resonator with zero built-in-tension is calculated to be $f_\text{out}=\SI{5.3}{\mega \hertz}$. However, if the same beam is actuated in-plane, the frequency is $f_\text{in}=\SI{45.7}{\giga \hertz}$, which shows that in-plane-actuated nanomechanical devices may operate at speeds consistent with modern optical communication systems.

\begin{figure}
\includegraphics[width=\columnwidth]{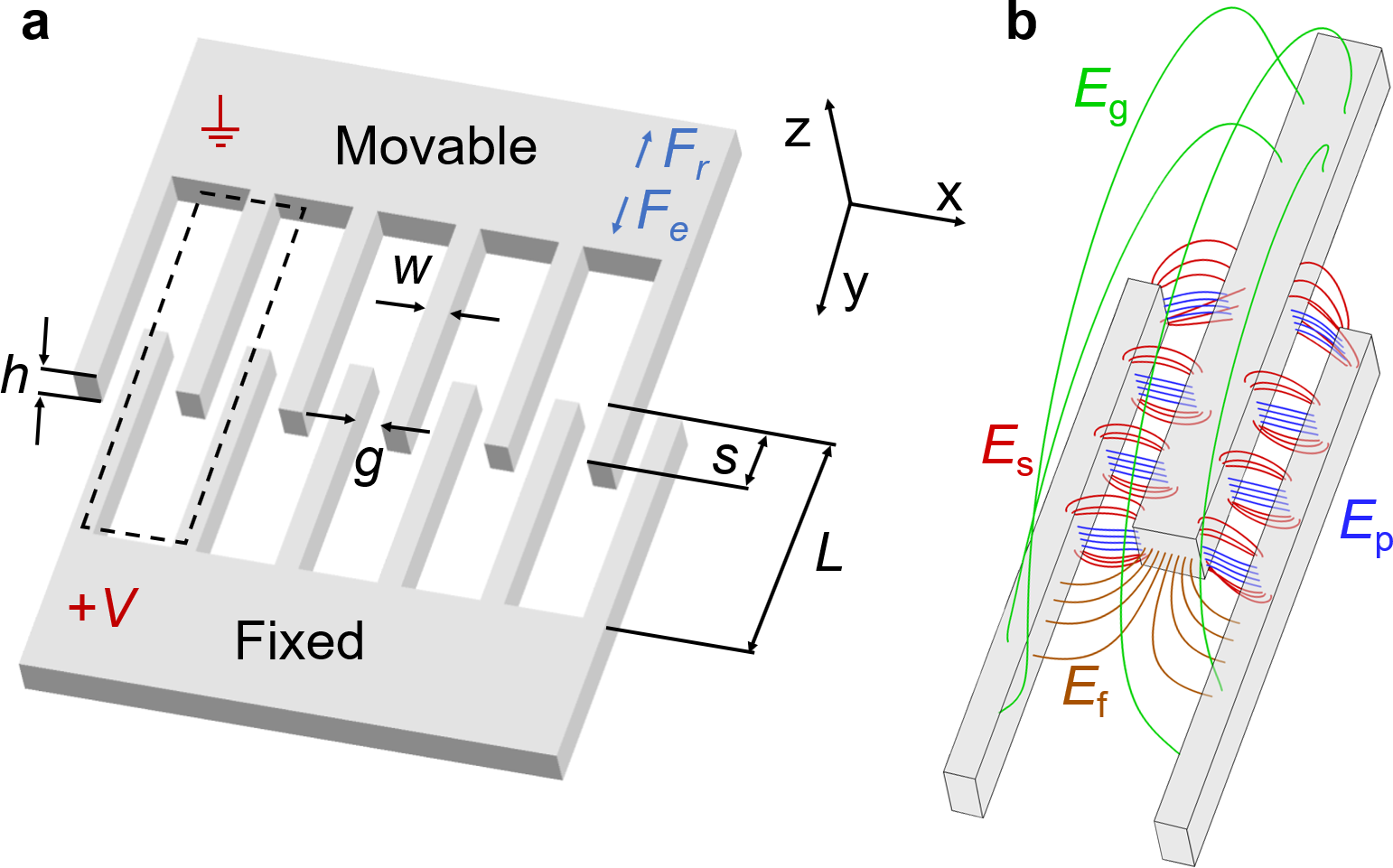}
\includegraphics[width=\columnwidth]{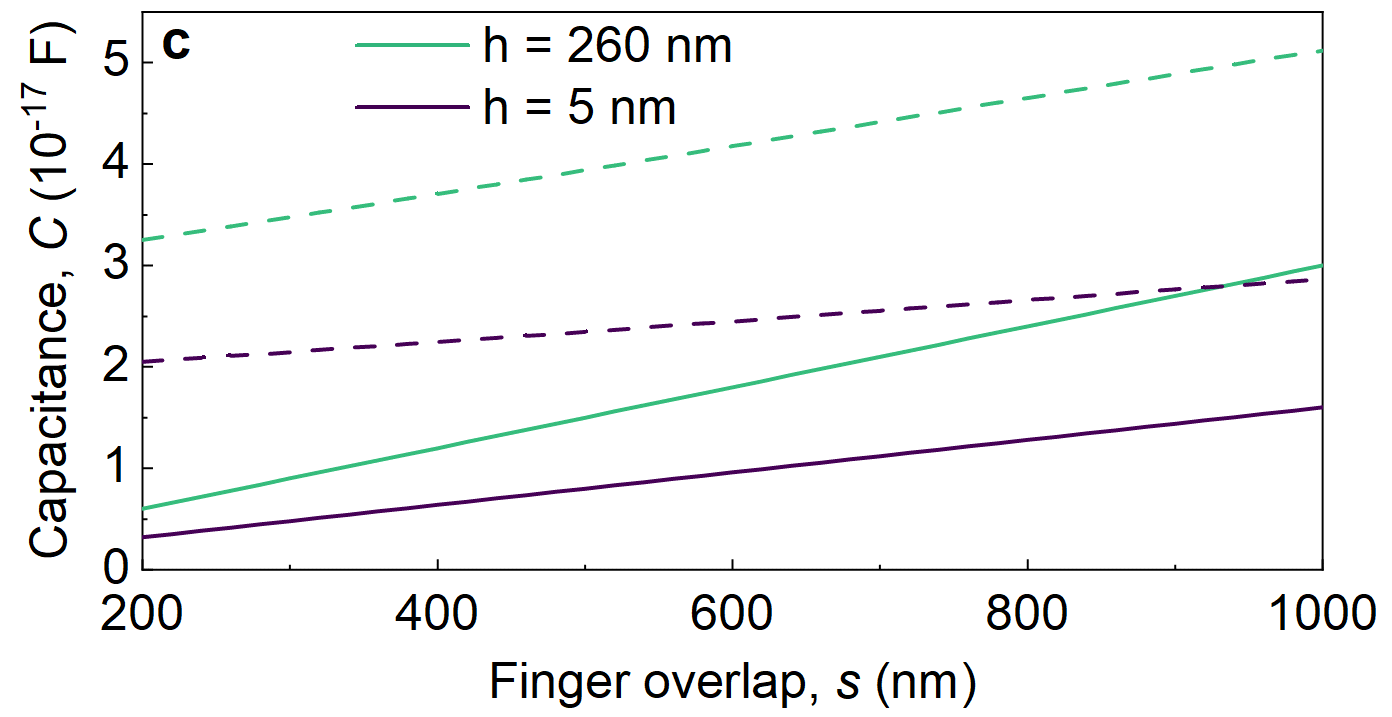}
\caption{\label{fig:stream2} Electric fields and capacitances of comb-drive actuators. (\textbf{a}) Illustration of a comb-drive actuator. (\textbf{b}) Selected field lines for a unit cell. (\textbf{c}) Capacitance of a unit cell as a function of finger overlap for a near-unity aspect ratio ($h=\SI{260}{\nm}$) as well as for nearly flat slabs ($h=\SI{5}{\nm}$). The full 3D capacitance is higher than the 2D approximation as it includes the extra capacitance of the tip fields $E_\text{f}$ and of the global fields $E_\text{g}$, but the line has a lower slope because this capacitance has a total negative contribution to the force. This capacitance does not change the linear behaviour of the actuator, that persists in the ultrathin membrane regime.}
\end{figure}

We finally turn to comb-drive actuators (Fig.~\ref{fig:stream2}\textbf{a}), which are commonly employed because of their linear capacitance-displacement relationship, enabling large displacements without a pull-in instability in the actuation direction \cite{Grade03,slender}. Traditionally, comb-drive actuators have been treated in the parallel-plate approximation, even in devices with near-unity aspect ratio \cite{vic1,vic2,vic3,vic4}. The electric field lines of a periodic cell in the finger array, shown in Fig.~\ref{fig:stream2}\textbf{b}, can be classified in three categories with distinct impact on the actuation force: fields created by the finger tips, $E_\text{f}$, fields between overlapping fingers, $E_\text{p}$ and $E_\text{s}$, and all remaining fields, $E_\text{g}$. The fields $E_\text{p}$ and $E_\text{s}$ correspond to the parallel-plate and co-planar-strip fields of the parallel beams. Their capacitance increases linearly with the finger overlap, $C =  C_\text{c}s$, where $ C_\text{c}$ is the 2D capacitance of the cross-section and $s$ is the overlap length, i.e., it gives rise to a constant force. The field $E_\text{f}$ encounters only a translational shift in the displacement direction without experiencing any redistribution in its profile and thus $\partial C_\text{f}/\partial s =0$, so it does not contribute to the force. The field lines of $E_\text{g}$ give rise to global forces with no counterpart in parallel plates, beams, and co-planar strips. They connect non-overlapping regions and terminate on top and/or bottom surfaces, i.e., they can only be understood in a 3D model. These fields are diminished by actuation due to the reduction of the non-overlapping region, leading to a negative contribution to the force, $\partial C_\text{g}/\partial s \leq 0$, which tends to disengage the fingers.
Previous work \cite{actPHYS} approximated the global force as $F_\text{g}=-4(w+g)\frac{\epsilon_0 V^2}{4 \pi s}$, which does not depend on the height and thus confirms the out-of-plane character. However, it also predicts a nonlinear dependence on the overlap but as we will show below, this nonlinearity vanishes for low aspect ratios.

\begin{table}[b]%The best place to locate the table environment is directly after its first reference in text
\caption{\label{tab:table1}%
Differential capacitance calculated by linear fitting of the capacitance simulations shown in Fig.~\ref{fig:stream2}\textbf{c}.
}
\begin{ruledtabular}
\begin{tabular}{lcc}

 \multicolumn{3}{c}{$\Delta C / \Delta y$ $(\SI{e-11  }{\farad \m^{-1}})$ } \\
 \\
\textrm{Approximation}&
\textrm{h = \SI{260}{\nm}}&
\textrm{h = \SI{5}{\nm}}\\
\colrule
 Parallel-Plate           & 1.3       & 0.025 \\
 Simulation 2D $(C_c)$    & 3.0       & 1.6  \\
 Simulation 3D            & 2.4       & 1.0 \\
\end{tabular}
\end{ruledtabular}
\end{table}

While the separation of field components is instructive for understanding the underlying physics of comb-drive actuators, the separation is not entirely rigorous and we therefore use a finite-element method to calculate the capacitance of a unit cell as a function of finger overlap, for fingers with width $w = \SI{250}{\nm}$, gap $g = \SI{350}{\nm}$, length $L = \SI{2}{\um}$, and two different heights of $h = \SI{260}{\nm}$ and $h = \SI{5}{\nm}$. We calculate the 2D cross-section capacitance $C_\text{c}$ and the full 3D capacitance $C$. The latter is compared to the approximation $C=C_\text{c} s$ in  Fig.~\ref{fig:stream2}\textbf{c}. We find that the full 3D capacitance is a linear function of the finger overlap, contrary to the global-force formula in Ref.~\onlinecite{actPHYS}. The contributions from the fingertip and the global fields ensure that the 3D capacitance is always higher, but its slope always lower, than the 2D approximation. To quantify the global forces we extract the 3D differential capacitance from a linear fit and compare it to the 2D simulation and the parallel-plate approximation, see Table \ref{tab:table1}. We assume that the difference in the 2D and the 3D calculations is caused by contribution of the global forces, which is quantified to be $\Delta C_\text{g}  / \Delta y = -\SI{0.6 e-11  }{\farad \m^{-1}}$ for both  $h = \SI{260}{\nm}$ and  $h = \SI{5}{\nm}$. This result suggests that the global forces take place between the outer surfaces of the fingers and do not depend on their height, in accordance with Ref.~\onlinecite{actPHYS}. These contributions significantly reduce the total electrostatic force by $20\%$ for $h = \SI{260}{\nm}$ and $38\%$ for $h = \SI{5}{\nm}$. Finally, we note that the parallel-plate approximation significantly underestimates the electrostatic force of actuators with an aspect ratio near unity and completely fails for ultra-small aspect ratios.

From the preceding discussions it is clear that the forces stemming from the finger width increase in importance with the reduction of height, impacting the miniaturization of the lateral dimensions of the comb-drive actuator. The width of a comb with $N$ finger pairs is $2N(w+g)$ and we are interested in choosing these parameters such that the actuation force is maximum for a given comb width. If the 3D effects are neglected, the figure of merit that must be maximised is $C_\text{c}/(g+w)$. In the limit of ultrathin actuators, we can approximate the cross-sectional capacitance as that of co-planar strips $C_\text{c}=C_\text{s}$. In this case, the force depends strongly on the ratio of width to gap and the force is maximum for $w \approx 0.4g$ (See Appendix) and as small as possible $g$. In the case of finite height, the fingers can be approximated as parallel beams with $C_\text{c}=C_\text{2D}$ given by Eq.~(\ref{eq:C2da}), which is maximized when both $g$ and $w$ are as small as possible. This is, however, not feasible because of an instability that collapses individual fingers with too small width, that was studied in the parallel-plate approximation in Ref~.\onlinecite{slender}. We linearize the contribution of the capacitance Eq.~(\ref{eq:C2da}) in the Euler-Bernoulli equation and we introduce the fringing field corrections to the model of Ref.~\onlinecite{slender} to obtain (See Appendix)
\begin{equation}
\label{eq:slender}
    \Tilde{V}^2 = V^2 \frac{\epsilon_0 L^4}{E w^3} \left( \frac{24}{g^3}-\frac{6.578 w^{0.23}}{h g^{2.23}} \right) \;.
\end{equation}
Equation (\ref{eq:slender}), along with the requirement that $\Tilde{V} \leq 3.516 $, puts bounds on the comb-drive parameters so that there is no instability. For example, for silicon with $V=\SI{30}{V}$, $g = \SI{350}{\nm}$, $h = \SI{260}{\nm}$ and $L = \SI{2}{\um}$, the minimum finger width is $w \geq \SI{35}{\nm}$. If the fringing field contribution is ignored and a parallel-plate model is used instead, the minimum width is $w \geq \SI{32}{\nm}$, showing that the fringing field only has a small effect on the instability as expected from its scaling.

Our findings show that contrary to prevalent assumptions, NEMS transducers with ultralow aspect ratios are not only electromechanically possible, they are in fact in many ways superior to their high-aspect-ratio counterparts. In such flat actuators, the fringing fields take over from the parallel fields, resulting in qualitatively similar transduction but with a reduced impact of pull-in instabilities. This surprisingly advantageous scaling to flat structures shows that a new realm of nanoelectromechanical experiments and devices are within experimental reach. For example, it enables electromechanical actuation at tens of gigahertz of 2D materials \cite{Bachtold08,Guttinger2017}, Floquet topological photonics \cite{Fang2012}, tunable optomechanical crystal resonators \cite{Pfeifer16} or quantum optomechanics based on superconducting resonators \cite{Teufel2011}. Finally, the fact that actuators with ultralow aspect ratios maintain a large capacitance has important implications for the scaling of nanoscale sensors, not only because the capacitance is larger than previously believed but more importantly because in-plane transduction offers ample opportunities for reducing and engineering the stray capacitance, which is the limiting factor in state-of-the-art NEMS capacitive sensors \cite{badscale}.

We gratefully acknowledge financial support from the Villum Foundation.

%We believe that electrostatic in-plane actuation will play an important role in the emerging field of reconfigurable metamaterials \cite{Ou2011}, where refractive index changes with a frequency in the order of  $\SI{20}{\giga \hertz}$ will enable optical topological Floquet insulators \cite{Fang2012}. We further envision stacking thin layers on top of each other and actuating them independently, leading to reconfigurable 3D metamaterials and $(1+1)$ topological photonic devices modulated in both space \cite{Rechtsman2013} and time.
\appendix
\section{Appendix}
\label{A1}

\subsection{Theory of electrostatic actuation}
%[mention cob drives and differential capacitance]
The most common actuation mechanism in micro- and nanoelectromechanical systems (MEMS and NEMS) is electrostatic actuation. It is based on the elastic deformation of solid bodies due to attractive electrostatic forces. In practice, this is achieved by applying a bias voltage, which can be tuned to control the deformation. The system forms a capacitor whose stored energy changes as it is deformed and is given by
\begin{equation}
    \label{eqs:capU}
    U = \frac{1}{2}CV^2,
\end{equation}
where $C$ is the capacitance of the system and $V$ is the potential difference. By applying the virtual-work principle, the electrostatic force is found to be \cite{grif}
\begin{equation}
    \label{eqs:ef}
    F_\text{e} = \frac{1}{2} \frac{\partial C}{\partial y} V^2,
\end{equation}
where $y$ is the geometric parameter that is changed.

We note that the naive calculation of the force by
\begin{equation}
    F = -\frac{\partial U}{\partial y}
\end{equation}
would give the wrong sign and predict electrostatic repulsion instead of attraction. It is only by considering the work done by a battery supplying the potential difference that the correct expression Eq.~(\ref{eqs:ef}) is obtained.

The deformation of the bodies results in a restoring force $F_r$. If the exact deformation is not of interest, the actuator can be described by a spring-mass model \cite{lump}, where its parts remain parallel. The spring depends only on the lumped parameter, $k$, that characterizes the mechanical properties of the system for a given displacement, $y$. In that case, the net force can be written as
\begin{equation}
\label{eqs:flump}
     F_\text{tot} = \frac{1}{2} \frac{\partial C}{\partial y} V^2 - ky \; .
\end{equation}
Stability at the point $y_0$ requires that $F_\text{tot}(y_0)=0$ and $\partial F_\text{tot}(y_0) /\partial y <0$. The stability conditions at the pull-in point, $y_\text{p}$, lead to 
\begin{align}
    y_\text{p} &= \frac{\partial C( y_\text{p})}{\partial y} \Big /\frac{\partial ^2 C( y_\text{p})}{\partial y^2} \; ,     \label{eqs:xpi} \\
    V_\text{p} &= \sqrt{2ky_\text{p} \Big / \frac{\partial  C( y_\text{p})}{\partial y}} \; .     \label{eqs:vpi} 
\end{align}
We note that in a mass-spring model, the pull-in distance is determined solely by the capacitance of the geometry. In the case where the capacitance is given by an inverse power law such as
\begin{equation}
\label{eqs:plaw}
    C = \frac{\lambda}{(1-y)^w}\; ,
\end{equation}
Eq.~(\ref{eqs:xpi}) reduces to
\begin{equation}
    y_\text{p}= \frac{1- y_\text{p}}{w+1}\; .
\end{equation}
%which is solved to give 
%\begin{equation}
%    y_\text{p}= \frac{1}{w+2}\; .
%\end{equation}
For example, the capacitance for a parallel-plate actuator is described by a model where $w=1$ and
\begin{equation}
   y_\text{p} = \frac{1}{3} \; .
\end{equation}
The distances in Eq.~(\ref{eqs:plaw}) are measured in units of the initial gap of the actuator and by restoring the units we find the maximum gap change,
\begin{equation}
    \Delta g= \frac{g_0}{3} \; .
\end{equation}

In other geometries, the electrostatic actuation can be significantly influenced by the deformation, which is the case for an actuator formed by two deformable beams. The coupled electrostatic-mechanical behaviour can be described by the Euler-Bernoulli equation, which, for a homogeneous beam, takes the form 
\begin{equation}
    \label{eq:Euler}
    E\frac{hw^3}{12}\frac{d^4D(x)}{dx^4}= \frac{1}{2} \frac{\partial C_\text{c}}{\partial g(x)} V^2
\end{equation}
where $E$ is Young's modulus and $h$, $w$ and $g$ are the height, width, and separation of the beams. The capacitance per unit length is the cross-section capacitance, $C_\text{c}$, and the deformation of an individual beam is $D(x)=\frac{g(x)-g_0}{2}$. Even though lumped spring-mass models can be useful, the inclusion of deformation in the analysis can drastically change the predicted actuator behaviour. For example, a spring-mass model was used in Ref.~\onlinecite{seeger} to predict that the pull-in instability of an electrostatic actuator can be completely avoided by a capacitive control scheme, which turned out to be a spurious effect when the deformations were taken into account \cite{Pelesko2001}.

\subsection{The capacitance of two co-planar rectangles}

\begin{figure}
\includegraphics[width=\columnwidth]{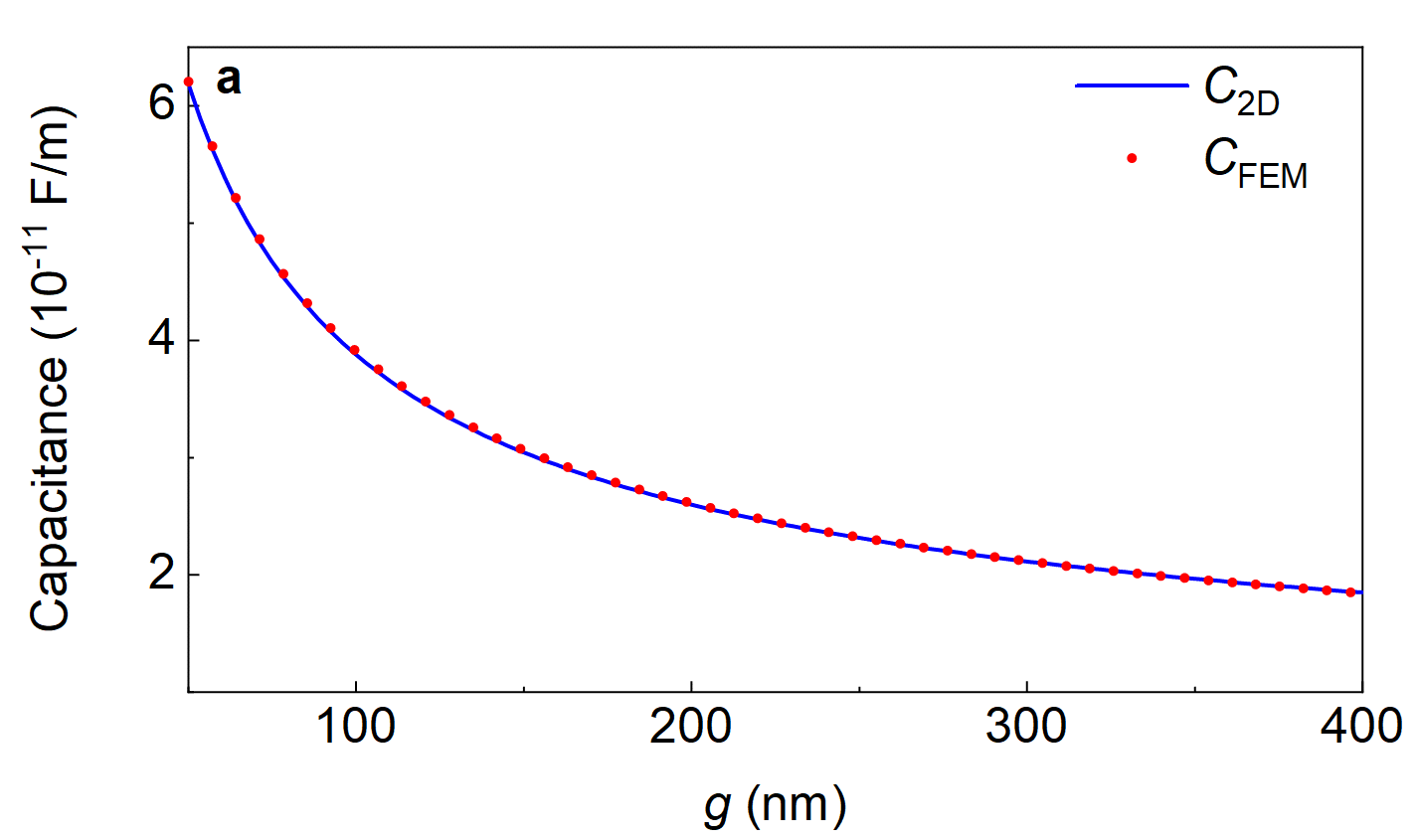}
\includegraphics[width=\columnwidth]{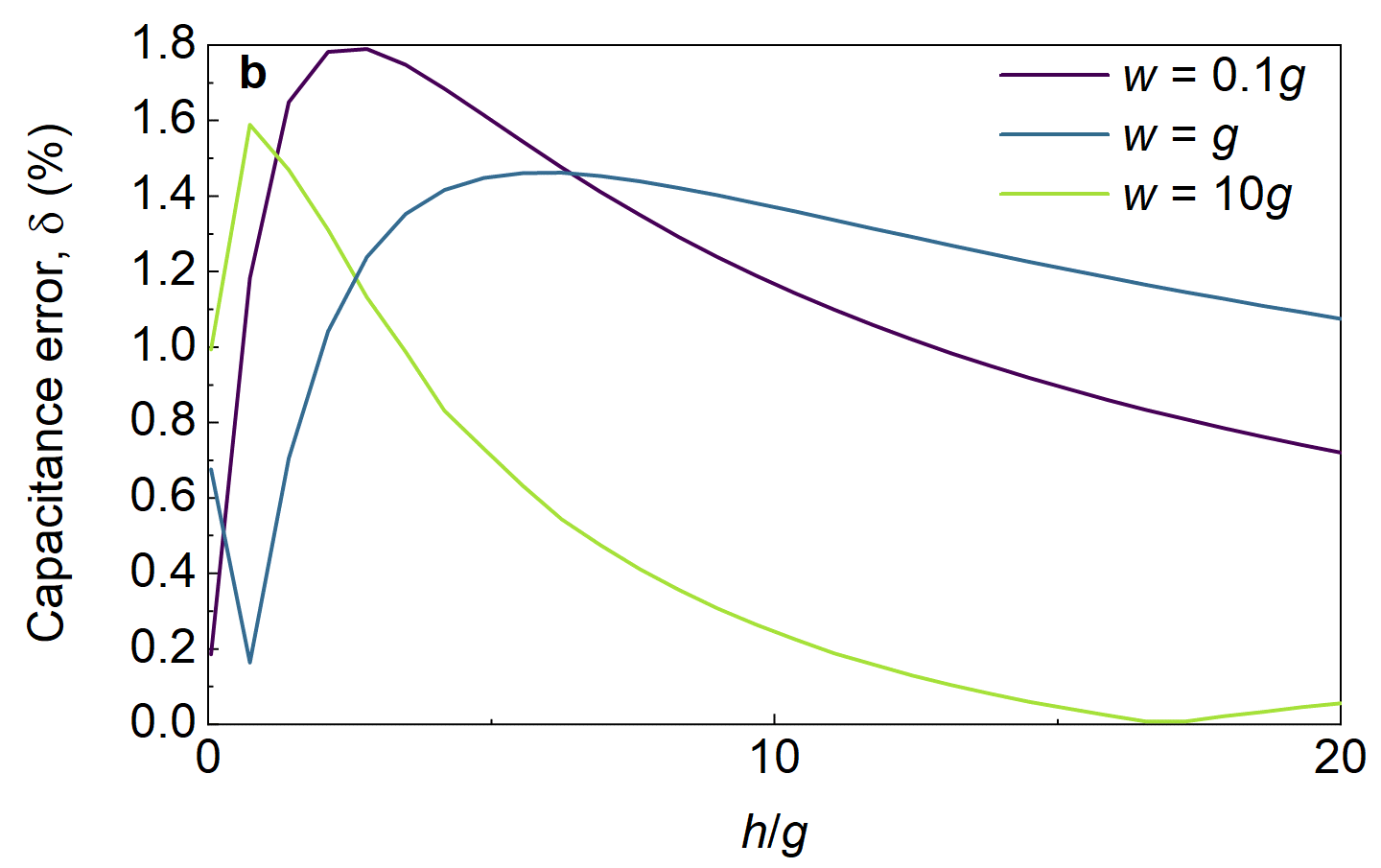}
\caption{\label{sfig:1} (\textbf{a}) Comparison of the empirical capacitance formula to the exact numerical calculation for the capacitance of two co-planar rectangles of $w= \SI{230}{\nm}$ and $h= \SI{220}{\nm}$. (\textbf{b}) Error of empirical formula for different ratios of height, width, and gap. Notably, the empirical formula remains accurate to within better than 2 percent even beyond the bounds for which it was derived.}
\end{figure}

There is no exact formula for the two-dimensional mutual capacitance, $C_\text{2D}$, of two co-planar rectangles but there is a plethora of approximations \cite{Leus04}. They are derived either from analytical techniques such as conformal mapping or fitting to finite-element-method (FEM) simulations. The analytical formulas are based on first principles but contain many terms with obscure physical meaning and they can only be derived for certain limits of geometrical ratios. The formulas obtained by FEM fitting are obviously heuristic but they benefit from clear physical interpretations. An empirical formula for a rectangle of height $h$ and width $w$, at a distance $g$ from a perfectly conducting wall, was derived and tested experimentally in Ref.~\onlinecite{Chuang_2012} who found,
\begin{equation}
\label{eqs:chuang}
    C = \epsilon_0 \left( 
    \frac{h}{g}+c \left( \frac{w}{g}  \right)^{a} + d \left( \frac{h}{w}  \right)^{b} + f
    \right) \; ,
\end{equation}
with $a=b=0.23$, $c = 3.31$, $d =0.73$, $f = -1.06$ and it is valid for $h/g > 0.05$ and $0.05<w/g<5$.
Since the capacitance of a conductor with its mirror image is half that of the conductor next to an infinite conducting plane \cite{capacitances} and the distance of an object to its mirror image is double that to the mirror plane, we determine the capacitance of two co-planar rectangles to be 
 \begin{equation}
\label{eqs:C2da}
   C_\text{2D} =  \frac{\epsilon_0}{2} \Bigg( \frac{2h}{g} + c \bigg( \frac{2w}{g} \bigg)^{a} + d \bigg( \frac{h}{w} \bigg) ^{b} +f \Bigg) \; .
\end{equation}
The first term of Eq.~(\ref{eqs:C2da}) corresponds to the capacitance of two parallel plates and the second term to the capacitance due to the fringing fields. The third term, which does not depend on the distance between the two conductors, is interpreted as coming from interaction between the top/bottom and rear sides or the individual rectangles. We compare the predicted capacitance per unit length $C_\text{2D}$ of Eq.~(\ref{eqs:C2da}) with the capacitance $C_\text{FEM}$ calculated by a FEM. Equation (\ref{eqs:chuang}) was derived \cite{Chuang_2012} by fitting to the capacitance calculated by the FEM software ANSYS and to facilitate a comparison, we therefore use another FEM software (COMSOL). Figure \ref{sfig:1}\textbf{a} shows the excellent agreement between $C_\text{2D}$ and $C_\text{FEM}$ for two rectangles of $w= \SI{230}{\nm}$ and $h= \SI{220}{\nm}$. Equation (\ref{eqs:C2da}) is further tested for three different values of the width-to-gap aspect ratio $w/g$ as a function of the height-to-gap aspect ratio $h/g$, where the error between the exact $C_\text{FEM}$ and $C_\text{2D}$ is quantified as 
 \begin{equation}
    \delta = \abs*{ \frac{C_\text{FEM}-C_\text{2D}}{C_\text{FEM}} } \; .
 \end{equation}
It can be seen from Fig.~\ref{sfig:1}\textbf{b} that the error is always below $2\%$ and that Eq.~(\ref{eqs:C2da}) gives good agreement even for $w=10g$, which is beyond the bounds of its formal validity.

\subsection{Electrostatics of co-planar strips}

The electric fields, $E_x$, $E_y$, and the capacitance per unit length, $C_\text{s}$, of two co-planar strips, with potential difference $V$, width $w$, and separation gap $g$, can be written in closed analytical form \cite{strips} as
\begin{align}
    E_x &= -\frac{V}{2K\left( q \right)} \text{Re} \left( f \left(z \right) \right) \; ,\\
    E_y &= \frac{V}{2K\left( q \right)} \text{Im} \left( f \left(z \right) \right) \; , \\
    C_\text{s} &= \frac{\epsilon_0 K \left( \sqrt{1-q^2} \right) }{K \left( q\right)}   \; ,  \\
\end{align}
with
\begin{align}
 q &= \frac{g}{g+2w} = \frac{1}{1+2\frac{w}{g}} \; , \\
    z &= x+iy \; , \\
    f \left(z \right)  &=   \frac{g}{\sqrt{ \left( g^2 -4k^2z^2 \right)  \left(g^2 -4z^2 \right) }}         \; ,  \\
\end{align}
 where
\begin{equation}
\label{eqs:kint}
    K \left(  k \right) = \int_0^{\pi/2} \frac{d \theta}{\sqrt{1-k^2 \sin^{2}\theta}} 
\end{equation}
is the complete elliptic integral of the first kind \cite{lawden89}.

\begin{figure}
    \includegraphics[width=\columnwidth]{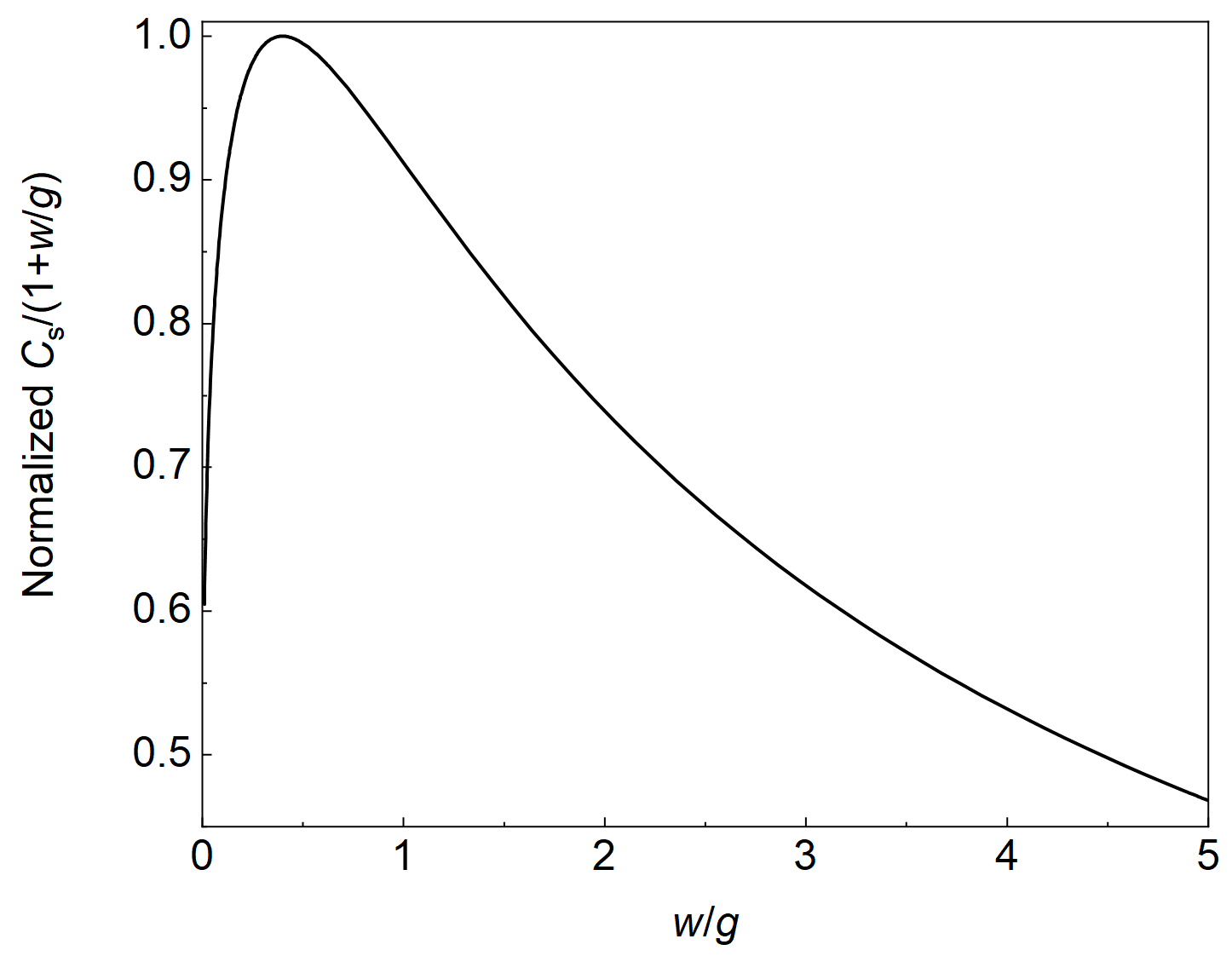}
    \caption{The figure of merit for achieving maximum force for a given comb gap as a function of width-to-gap ratio. This quantity varies significantly with $w$ and reaches its maximum value for $w/g = 0.398$.}
    \label{figs:maxwg}
\end{figure}

The capacitance of a comb-drive actuator whose fingers are thin enough to be considered co-planar strips is given by $C=N C_\text{s} s$, with $s$ being the finger overlap and $N$ the number of finger pairs. Ignoring the global-force contribution, the electrostatic force is given by
\begin{equation}
    F_\text{e} = \frac{\partial C}{\partial s} = N C_\text{s}\;.
\end{equation}
Since the comb has a width of $2N(w+g)$, the electrostatic force per unit of width is 
\begin{equation}
F = \frac{C_\text{s}}{2(w+g)} = \frac{1}{2g} \frac{C_\text{s}}{1+\frac{w}{g}} \;.
\end{equation}
In order to maximize the force for a given gap $g$, we need to maximize the function $C_\text{s}/\left( 1+ w/g \right)$, which is shown in Fig.~\ref{figs:maxwg}. It changes significantly as a function of width to gap ratio, $w/g$, which directly impacts the prospects for scaling to ultrathin actuators. We numerically determine that its maximum value is achieved for $w/g = 0.398$ and $g$ as small as possible. 

\subsection{Asymptotic behaviour of the pull-in displacement}

\begin{figure}
    \includegraphics[width=\columnwidth]{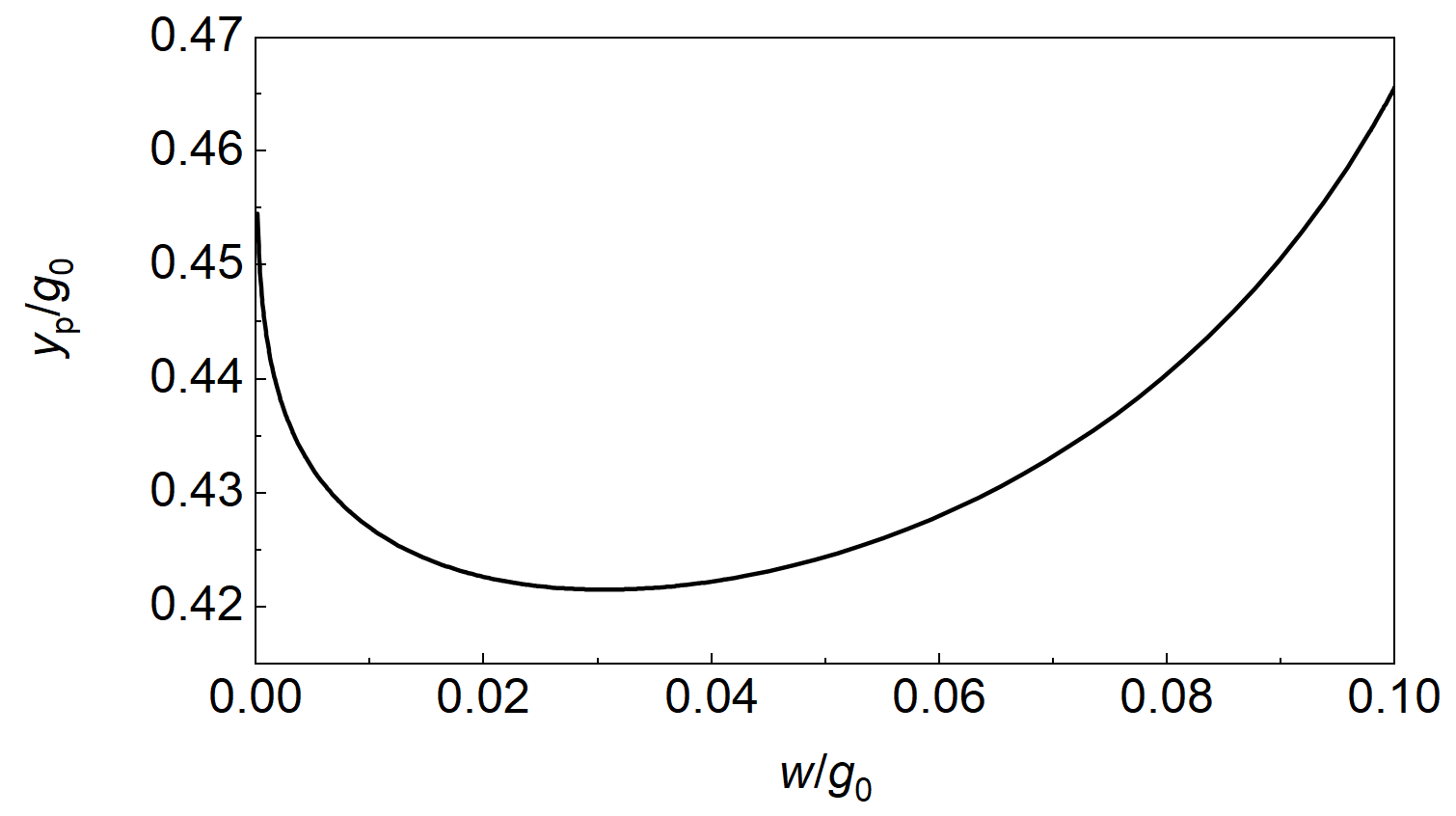}
    \caption{Normalised pull-in displacement as a function of normalised strip width for the asymptotic expression for a co-planar-strip actuator. The non-monotonic behaviour of the pull-in displacement is explained by the non-trivial dependence of the capacitance on the finger width.}
    \label{figs:yp}
\end{figure}

The limiting behaviour of the co-planar strip actuator can be analysed by using asymptotic expressions for the elliptic integral, which can be derived by Taylor expansion of Eq.~(\ref{eqs:kint}).

\subsubsection{The $w \to \infty$ limit}
The capacitance per unit length, $C_\text{s}$, of a co-planar strip actuator with a strip width much larger than its gap is given by the asymptotic expansion
\begin{equation}
\label{eqs:csinfty}
    \frac{C_s}{\epsilon_0} \sim \frac{6 \ln 2 }{\pi} + \frac{2}{\pi} \ln  \left( \frac{w}{g} \right) + \cdots  \left( w \to \infty \right) \; .
\end{equation}
Setting $g=g_0-y$ and inserting in Eq.~(\ref{eqs:xpi}), we obtain
\begin{equation}
    y_\text{p}=g_0-y_\text{p} \; ,
\end{equation}
which shows that the maximum actuation range for a co-planar strip asymptotically approaches
\begin{equation}
    y_\text{p}=\frac{g_0}{2} \; .
\end{equation}

\subsubsection{The $w \to 0$ limit}
The asymptotic expansion for the case of very small width-to-gap ratios is calculated to be
\begin{equation}
\label{eqs:cs0}
    \frac{C_s}{\epsilon_0} \sim \frac{\pi}{ \ln  \left( \frac{4g}{w} \right)}- \pi \frac{ w}{g} \frac{1}{ \ln  \left( \frac{4g}{w} \right)^2}  + \cdots  \left( w \to 0 \right) \; .
\end{equation}
In this limit, the strips become point-like and can be approximated as infinitely thin and long wires. In fact, we expect every 2D capacitor to reach the same limit as the lateral dimensions of the constituent bodies shrink to zero. For example, the capacitance per unit length of an actuator formed by two infinitely long cylinders is given by \cite{jackson99}
%\begin{multline}
\begin{align}
\label{eqs:cl}
    \frac{C_l}{\epsilon_0} &=  \frac{\pi}{	\operatorname{arcosh}\left( \frac{g+2R}{2R} \right) } \\
    &\sim  \frac{\pi}{ \ln  \left( \frac{2g}{R} \right)}- \pi \frac{ 2R}{g} \frac{1}{ \ln  \left( \frac{2g}{R} \right)^2}  + \cdots  \left( R \to 0 \right) \; .
%\end{multline}
\end{align}
It is clear from Eqs.~(\ref{eqs:cs0}) and (\ref{eqs:cl}) that in this limit, the co-planar strips have the same capacitance as two cylinders with radius $R=w/2$. Substituting $g = g_0 -y$ and using Eq.~(\ref{eqs:xpi}) we obtain the pull-in displacement curve shown in Fig.~\ref{figs:yp}. The asymptotic expansion captures the non-monotonic behaviour of the pull-in instability but higher order terms are needed to approach the exact point for different geometries. We do conjecture, however, that this behaviour is common in all 2D electrostatic actuators.

\subsection{Instability of individual comb-drive-actuator fingers}

\begin{figure}
\includegraphics[width=\columnwidth]{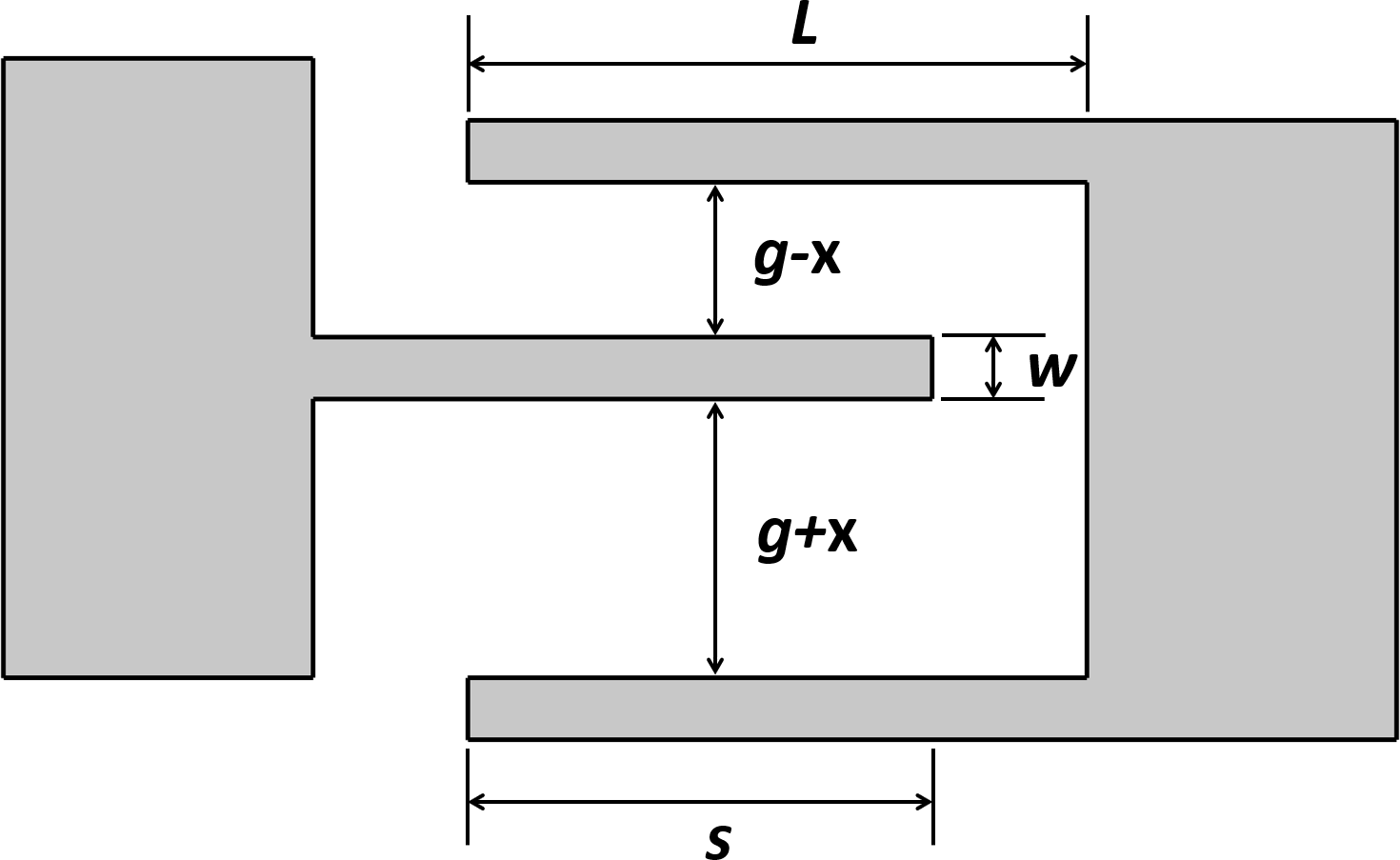}
\caption{\label{fig:finger} A comb-drive actuator finger that has been slightly displaced by $x$. If the electrostatic forces exceed the elastic restoring-force of the finger, it will bend and collapse onto the nearest finger. 
}
\end{figure}
We calculate the side-pull-in instability of an individual comb-drive finger by solving the Euler-Bernoulli equations with a capacitance that includes the fringing-field contributions. If we ignore the global forces, we can approximate the capacitance per unit length of the individual finger (Fig.~\ref{fig:finger}) displaced by $x$ as $C = C_\text{2D} \left( g + x \right) + C_\text{2D} \left( g - x \right) $. Since we are interested in the point where a small perturbation would make the finger unstable, we can consider the displacement $x$ to be very small and expand the Euler-Bernoulli equation to first order to obtain for fingers of length $L$, overlap $s$, width $w$, height $h$ and gap $g$
\begin{equation}
   E\frac{hw^3}{12}\frac{d^4 x}{ds^4}= \epsilon_0  V^2  \left( \frac{2 h}{g^3} + \frac{2^{a-1} a (a+1) c \left(\frac{w}{g}\right)^a}{g^2} \right) x  \; ,
\end{equation}
which can be brought to the dimensionless form 
\begin{equation}
\label{eqs:slender}
    \frac{d^4 \tilde{x}}{d \tilde{s} ^4} = \tilde{V}^2 \tilde{x}  \; ,
\end{equation}
with $\tilde{x} = x/g$, $\tilde{s}= s/L$ and
\begin{equation}
    \tilde{V}^2 =  \frac{ 12 \epsilon_0 L^4}{E h w^3} \left( \frac{2 h}{g^3} + \frac{2^{a-1} a (a+1) c \left(\frac{w}{g}\right)^a}{g^2} \right)  V^2 \; .
\end{equation}

The solutions of Eq.~(\ref{eqs:slender}) were studied in Ref.~\onlinecite{slender} in the context of the parallel-plate approximation. It was found that a reqruirement for Eq.~(\ref{eqs:slender}) to have bound solutions is $\tilde{V} \leq 3.516$. This result can be immediately carried over to the case where the fringing fields are taken into account to derive an inequality that all the parameters of the problem must obey,
\begin{equation}
\label{eqs:finger}
    \frac{ 12 \epsilon_0 L^4}{E h w^3} \left( \frac{2 h}{g^3} + \frac{2^{a-1} a (a+1) c \left(\frac{w}{g}\right)^a}{g^2} \right)  V^2  \leq 12.36  \; .
\end{equation}
The inequality (\ref{eqs:finger}) shows that the comb-drive parameters cannot be chosen independently. Some of these parameters will be constrained by physics and others by fabrication. For example, the voltage $V$ must be below the value of breakdown due to field emission, the Young's modulus $E$ is a property of the material and the device height $h$ is constrained by applications. It is also desirable to choose $w \approx 0.4g$ as shown in Fig.~(\ref{figs:maxwg}) and it is also constrained by fabrication limitations. Finally, this leaves the comb gap $g$ and the finger length $L$ to be chosen so that the pull-in instability is avoided.

\providecommand{\noopsort}[1]{}\providecommand{\singleletter}[1]{#1}%\providecommand{\noopsort}[1]{}\providecommand{\singleletter}[1]{#1}%

\bibliographystyle{apsrev4-2}
%\bibliography{actbib}% Produces the bibliography via BibTeX.

\end{document}